\documentclass[5p,times,authoryear]{elsarticle}

\usepackage{ecrc}

\volume{00}

\firstpage{1}

\journalname{Astronomy and Computing}

\runauth{O. Creaner, J. Walsh, K. Nolan \& E. Hickey  }

\jid{astron.comput.}

\jnltitlelogo{Astronomy and Computing}

\usepackage{amssymb}

\usepackage[figuresright]{rotating}

\usepackage[table]{xcolor}

\usepackage{nameref}

\usepackage{hyperref}

\begin{document}

\begin{frontmatter}



\dochead{}

\title{The Locus Algorithm IV: Performance metrics of a grid computing system used to create catalogues of optimised pointings}


\author[ITTD,DIAS,LBNL]{Ois\'{i}n Creaner\corref{cor1}}
\cortext[cor1]{Corresponding author}
\ead{creanero@gmail.com, oocreaner@lbl.gov}

\author[GI]{John Walsh}

\author[ITTD]{Kevin Nolan}
\ead{kevin.nolan@tudublin.ie}
\author[ITTD]{Eugene Hickey}
\ead{eugene.hickey@tudublin.ie}

\address[ITTD]{Technological University Dublin, Tallaght Campus, Dublin 24, Ireland}
\address[DIAS]{Dublin Institute for Advanced Studies, 31 Fitzwilliam Place, Dublin 2, Ireland}
\address[GI]{Grid-Ireland, School of Computer Science and Statistics, Trinity College, Dublin 2, Ireland}
\address[LBNL]{Lawrence Berkeley National Laboratory, 1 Cyclotron Road, Berkeley, California, USA}

\begin{abstract}
This paper discusses the requirements for and performance metrics of the the Grid Computing system used to implement the Locus Algorithm to identify optimum pointings for differential photometry of 61,662,376 stars and 23,779 quasars. 
Initial operational tests indicated a need for a software system to analyse the data and a High Performance Computing system to run that software in a scalable manner.  Practical assessments of the performance of the software in a serial computing environment were used to provide a benchmark against which the performance metrics of the HPC solution could be compared, as well as to indicate any bottlenecks in performance.
These performance metrics indicated a distinct split in the performance dictated more by differences in the input data than by differences in the design of the systems used.  This indicates a need for experimental analysis of system performance, and suggests that algorithmic complexity analyses may lead to incorrect or naive conclusions, especially in systems with high data I/O overhead such as grid computing.  Further, it implies that systems which reduce or eliminate this bottleneck such as in-memory processing could lead to a substantial increase in performance.

\end{abstract}


\end{frontmatter}


\section{Introduction}
\label{Introduction}
The Locus Algorithm, as proposed by \citet{creaner2010large} and detailed in \citet{locuspaper} is an algorithm to calculate the optimum pointing for differential photometry for a given target by adjusting the central point of the field of view (FoV) of a given telescope such that the target remains in the FoV and the maximum number and quality of reference stars are included in the FoV with it.  A software system which harnesses this algorithm was developed as shown in \citet{creaner2016thesis} and \citet{locus_software_paper} and used with 40,000 quasars from the Sloan Digital Sky Survey (SDSS) catalogue \cite{schneider2007sloan} as input targets to generate pointings for 23,779 quasars as discussed in \citet{quasarpaper,ZenodoQuasarCatalogue} and with 357,175,411 point sources from SDSS \cite{abazajian2009seventh} as inputs to produce pointings for 61,662,376 stars for use as candidates for exoplanet observation as shown in \citet{ZenodoXOPCatalogue}.  The software is available on GitHub \cite{githubrepo}.

Operationally, a single instance of the algorithm pipeline was measured to have a run time of 0.25-1.0 seconds using a commercial off-the-shelf PC.  Assuming linear scaling of runtime, this leads to a predicted an overall runtime of between 2.8 and 11 years for the generation of a catalogue of exoplanet candidates from SDSS.  This demonstrated a strong requirement for a High Performance Computing (HPC) solution which would reduce this runtime to a practical level.  
These initial performance metrics also characterised the pipeline as being I/O limited, in that the runtime was dominated by data I/O operations over algorithmic processing time.

These characteristics are expected to become more significant in analyses of future catalogues, such as those produced by Gaia (which by its second data release has already exceeded SDSS's fifteeth in number of detected objects) \cite{brown2018gaia,aguado2019fifteenth} or LSST, which is expected to eventually produce data for more targets than were used in this project by two orders of magnitude \cite{ivezic2019lsst,abazajian2009seventh}.

For this project, it is conceptually possible that the analysis of each target can be carried out independently, lending itself well to parallelisation.  Real systems, however, can be subject to bottlenecks or other limitations, some of which can be difficult or impossible to anticipate without testing. \cite{ueda2009challenge}  This paper discusses the metrics observed by practical implementation of this system.  The system is described briefly in Section \ref{System} and in more detail in \citet{locus_software_paper}.

\section{Computing System and Infrastructure}
\label{System}
\begin{figure}[!htb]
        \center{\includegraphics[width=.47\textwidth]
        {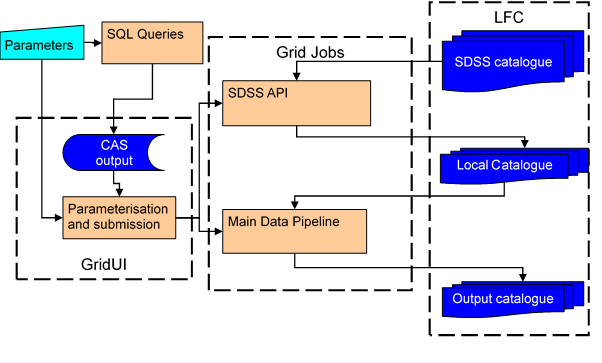}}
        \caption{\label{fig:grid_jobs} Conceptual structure of grid jobs as used in this project.  Orange rectangles represent programs.  Blue shapes represent catalogues or data files.  Cyan shapes represent console input.  Dashed rectangles represent how the interface, grid processes and data are separated in the grid system.  Copied from \citet{creaner2016thesis}.}
      \end{figure}

The objective of this project was to produce a system capable of analysing large astronomical catalogues to identify optimum pointings for differential photometry of a variety of classes of objects.  Figure \ref{fig:grid_jobs} illustrates the basic architecture of the system.  A more detailed discussion on the software system is published in \citet{locus_software_paper}.

While SDSS was selected as the input catalogue for the first iteration of this system, flexibility to adapt to novel cases was a core design objective \cite{creaner2016thesis}.  As a result, a data abstraction layer was required, whereby data from the source catalogue (in this case SDSS) is passed through and Application Programming Interface (API) to create data in a format that is source-independent, known as the Local Catalogue.  This allows for a data processing pipeline to be created that carries out the main task and produces the Output Catalogues.

User input is supplied through a command line interface.  By choosing particular parameters and providing paths to the input data for each of the API and the Pipeline, the user can select which particular analyses are carried out and thus determine the outputs e.g. the Quasar Catalogue \cite{ZenodoQuasarCatalogue,quasarpaper} or Exoplanet Target Catalogue \cite{ZenodoXOPCatalogue}.

The \texttt{glite} grid system separates user input, grid processing, and data storage into separate logical elements \cite{glite}.  User input is submitted through a system called gridUI, a computer with similar specifications to the grid's worker nodes with the ability to submit grid jobs and request data from the Logical File Catalogue (LFC) \cite{coghlan2005grid}.  Grid jobs are defined in .jdl format files which define an executable which carries out the task required on the grid's compute nodes \cite{glite}.  

Data files in the grid is accessed through the LFC.  The LFC is a distributed file system in which the user or grid jobs can upload data through \texttt{glite} commands.  Uploaded data may be stored physically on any of the grid sites, and given a name which mimics a UNIX file system name \cite{glite}.  This data can then be accessed by authorised users or jobs by further \texttt{glite} commands to download the data to the local device's filesystem (i.e. gridUI or the Worker Node) \cite{glite}.
      
\section{Metrics}
\label{Metrics}
This section presents the characteristics of the data in Subsection \ref{DataMetrics}, the nature of the processing challenge in Subsection \ref{ProcMetrics} and the performance of the resulting grid solution in Subsection \ref{GridMetrics}.  These metrics are presented and explained briefly here, but their interactions and the implications for the design of future projects is outlined in the \nameref{Discussion}, Section \ref{Discussion}.

\subsection{Data Metrics}
\label{DataMetrics}     

As discussed above, this project centred on an I/O limited paradigm, where each individual calculation was relatively straightforward, but scaling these calculations to many millions of iterations, and repeated data access led to a demand for a HPC solution.  Therefore, it is essential to understand the nature and structure of the data used in this project to understand the project as a whole.

\begin{center}
\begin{table*}[htb!]
\begin{tabular}{ |p{3cm}|p{2.5cm}|p{2.5cm}|p{2.5cm}|p{2.5cm}| } 
\hline
{\bfseries } & {\bfseries SDSS DR7 Catalogue\cite{abazajian2009seventh}} & {\bfseries Local Catalogue} & {\bfseries Quasar Catalogue\cite{ZenodoQuasarCatalogue,quasarpaper}}& {\bfseries Exoplanet Catalogue\cite{ZenodoXOPCatalogue}}\\ \hline
\hline
Entries & 357,175,411 & \textasciitilde 86,000,000 & 40,000. & 67,043,579 \\ \hline
Files & 421,388 & 358,076 & 40. & 1,598 \\ \hline
Directories & 3,290. & 2,609 & 1 & 7\\ \hline
Size on disk & 4.76 TB & 6.89 GB & 3.43 MB & 5.02 GB\\ \hline
Mean Size per file & 11.8 MB & 20.2 kB & 87.8 kB & 3.22 MB\\ \hline
Mean Size per entry & 14.6 kB & \textasciitilde 86 B & 87.8 B & 80.3 B\\ \hline
Mean Entries per file & 847 & 240 & 1,000. & 41,955\\ \hline

 \end{tabular} 
 \caption{Data metrics for the various datasets used and created during this project.  Approximate or rounded values are indicated with a tilde (\textasciitilde), while values with significant trailing zeroes are indicated with a decimal point (.). The Mean Size per entry is calculated by dividing the total size of the catalogue by the number of entries and as such includes contributions from header data spread per entry.  Copied from \citet{creaner2016thesis}}
 \label{table:datametrics} 
\end{table*}
\end{center}
All data used in this project was stored in the Flexible Image Transport System (FITS/.fit) file format.  FITS files consist of one or more Header Data Units, each of which includes a header which describes the data and the data unit which contains the data stored as a variety of binary data types.  These data types include both atomic data types such as floating point numbers and compound data types such as vectors containing multiple elements.  For this project, FITS data tables were used, with each row representing a detected object in the SDSS catalogue, and each column recording a different data variable about those objects.  

\subsubsection{Source Catalogue}
\label{Source}
As indicated in Section \ref{System}, the source data in this project was the SDSS DR7 catalogue.  This catalogue contained many records which did not meet the SDSS "Clean sample of point sources" criteria which is used to exclude duplicates, non-point sources such as galaxies and other targets in the SDSS catalogue that are not suited as candidate reference stars \cite{SDSScleansample}.  In addition, the source catalogue contained more variables than were required for the algorithm.  In order to improve the performance of the system, the Local Catalogue was generated which only included the required data.  
\subsubsection{Local Catalogue}
\label{Local}
In doing so, as shown in Table \ref{table:datametrics}, the number of entries in the catalogue was cut down to 24\% of the original catalogue entries.  In addition the data per entry was reduced from 150 columns to just 7, reducing the data required per entry to 0.59\% of what was required initially.  The total reduction in the data volume was to 0.14\% of the initial data.  This facilitated data access and transfer during the generation of Output Catalogues.  The mean size per entry is larger than might be predicted by a na\"{i}ve calculation because, for files containing few records, the contribution to file size from the header becomes significant.

\subsubsection{Output Catalogues}
\label{Outputs}
There were two Output Catalogues generated using the main software pipeline.  The first was the quasar catalogue \cite{ZenodoQuasarCatalogue,quasarpaper}. This catalogue used  the SDSS 4\textsuperscript{th} Quasar Catalogue \cite{schneider2007sloan} to provide a target list.  Of those targets, a sample of 40,000 were processed and pointings produced for 23,779 of them \cite{quasarpaper}.  The generation of this catalogue, while not computationally intensive enough to absolutely require a HPC solution in its own right, nevertheless served as a useful test of the grid implementation.  The other Output Catalogue was the Exoplanet Catalogue \cite{ZenodoXOPCatalogue}.  In this, every one of the \textasciitilde 86,000,000 objects in the Local Catalogue was used as both a target and a candidate reference star.  A total of 67,043,579 were successfully analysed generating optimum pointings for differential photometry of 61,662,376 stars \cite{ZenodoXOPCatalogue}.  As a result, this operation \textbf{did} mandate a HPC solution.

The Output Catalogues have the same data structure, but a different size per entry.  As in Subsection \ref{Local}, this is explained by the contribution of the Header unit to the overall file size.  As the number of entries per file trends upwards, the contribution from the header unit trends towards zero.  The Quasar Catalogue has been published in \citet{ZenodoQuasarCatalogue} and described in \citet{quasarpaper} and the Exoplanet Catalogue has been published in \citet{ZenodoXOPCatalogue} and a paper describing it is in preparation.

\subsection{Processing Metrics}
\label{ProcMetrics}

\begin{table*}[htb!]
\begin{tabular}{|p{3.6cm}|p{3.6cm}|p{3.6cm}|p{3.6cm}|}
\hline
                                                                                        & \textbf{API (Local Catalogue)} & \textbf{Pipeline (Quasar Catalogue)} & \textbf{Pipeline (Exoplanet Catalogue)} \\ \hline \hline
\textbf{Unit of Work}                                                                   & Source Catalogue file          & Quasar                               & Field                                   \\ \hline 
\textbf{No. of Units}                                                                   & 421,388                        & 77,429                               & 358,076                                 \\ \hline 
\textbf{Processing Time per unit}                                                       & 0.02887 s                      & 0.15 s                               & 36.0 s                                  \\ \hline 
\textbf{Average Input per unit}                                                         & 1 Source Catalogue file        & 11.73 Local Catalogue files          & 20.32 Local Catalogue files             \\ \hline 
\textbf{Data in per unit}                                                               & 11.8 MB                        & 237.1 kB                             & 410.5 kB                                \\ \hline 
\textbf{Access time per unit}                                                           & 4.5 s                          & 23.5 s                               & 40.6 s                                  \\ \hline 
\textbf{Mean output per unit}                                                           & 1 Local Catalogue file         & 1 output file entry                  & 240 output entries                      \\ \hline 
\textbf{Data out per unit}                                                              & 20.2 kB                        & 90 B                                 & 18.75 kB                                \\ \hline 
\textbf{Ratio of Access to Processing time} & 156:1                     & 157:1                            & 1.12:1                                  \\ \hline 
\end{tabular}

 \caption{ Metrics on processing time and data used per work unit on each of the three main grid jobs.  Copied from \citet{creaner2016thesis}}
 \label{table:processingmetrics} 
\end{table*}

As shown in Section \ref{System}, the production of the catalogues consisted of two major steps, the API and the Pipeline.  The API generated the Local Catalogue, a substantially reduced version of the SDSS Catalogue designed to be used as an input for Pipeline jobs. The pipeline took in a list of targets and accessed the Local Catalogue to generate Output Catalogues of targets with their respective pointings. 

\subsubsection{API Processing}
\label{APIProc}
The API worked by extracting relevant data from the Source Catalogue (i.e. SDSS) to produce an intermediate data format called the Local Catalogue.  For each file in the SDSS Catalogue, one corresponding file was produced in the Local Catalogue.  In this process, which consisted of filtering the SDSS Catalogue to remove rows pertaining to non-point sources and columns which were not needed for the pipeline, I/O operations, in particular network I/O to the LFC dominated by a factor of approximately 156:1 as illustrated in Table \ref{table:processingmetrics}.
\subsubsection{Pipeline Processing}
\label{PipProc}
The pipeline is the central element of the implementation of the Locus Algorithm.  It is given targets and parameters by the user, and processes these targets together with data relating to candidate reference stars to produce the optimum pointing for each target.  The data from these targets are compiled into an output file, and a collection of these output files forms the Output Catalogue. There were two modes of operation for the pipeline.  Those modes are the Target List mode and the Catalogue mode \cite{locus_software_paper}.
\paragraph{Target List Mode}
\label{TargetList}
In the first mode of operation, exemplified by the Quasar Catalogue job, \cite{quasarpaper} the pipeline was provided with a list of individual target objects to consider.  For each of these targets individually, the Local Catalogue files containing candidate reference stars were identified and loaded into memory forming a mosaic, the algorithm was run, the pointing for that target identified and written to the output file and then the process restarted.  This meant that each mosaic of Local Catalogue files was used only once per target.  As a result, network I/O again dominated over operational processing time.
\paragraph{Catalogue Mode}
\label{CatalogueList}
The other mode of operation is to use the Local Catalogue itself as the target list, as exemplified by the Exoplanet Target Catalogue job \cite{ZenodoXOPCatalogue}.  In this mode, each Local Catalogue file is used as a target, and a mosaic is generated consisting of the neighbouring Local Catalogue files in which any star within that neighbouring file could be included in a FoV with a star in the target field.  For each star within the target file, the mosaic is retained, and used to identify candidate references for that star.  The algorithm then operates on that star and its candidate reference stars as discussed in \citet{locuspaper}.  The mean number of valid targets in a field is 240, which means that typically 240 instances of the algorithm are run for each mosaic that is generated.  This substantially reduces the ratio of I/O operations to processing operations.  In the case of the Exoplanet Catalogue job, the ratio of I/O time to Processing time was 1.12:1.  Note that confounding variables such as a change in FoV size between the Exoplanet and Quasar Catalogues means the I/O:Processing ratio between the two jobs cannot be directly compared.

\subsection{Grid Metrics}
\label{GridMetrics}

A major driver towards using a HPC solution is the scale up factor of speed when compared with a serial computing solution: that is how many times faster the job is executed when compared with that job executed in a serial environment.  Implementation of a HPC solution will typically require some overhead, in the form of job management or unavoidably serial components, which may include access to data on a central repository if not appropriately managed.  As a result, the scale-up factor is typically lower than the number of nodes used in the execution.  The ratio between the scale up factor and the number of nodes used is the HPC Efficiency.

\begin{table*}[htb!]
\begin{tabular}{|p{4cm}|p{3.5cm}|p{3.5cm}|p{3.5cm}|}
\hline
                                   & \textbf{Local Catalogue Job} & \textbf{Quasar Job} & \textbf{Exoplanet Job} \\ \hline \hline
\textbf{Work Units per job}        & 1,000                        & 1,000               & 200                    \\ \hline 
\textbf{Input data per job}        & 11.5 GB                      & 231 MB              & 80.2 MB                \\ \hline 
\textbf{No. of jobs}               & 422                          & 78                  & 1,791                  \\ \hline 
\textbf{No. successful jobs}       & 359                          & 40                  & 1,598                  \\ \hline 
\textbf{Percentage successful}     & 85.10\%                      & 51.30\%             & 89.20\%                \\ \hline 
\textbf{Output data per job}       & 19.7 MB                      & 85 kB               & 3.22 MB                \\ \hline 
\textbf{Serial time per job}       & 1.25 hours                   & 6.57 hours          & 4.25 hours             \\ \hline 
\textbf{Serial overall time}       & 21.9 days                    & 21.2 days           & 317 days               \\ \hline 
\textbf{Nodes used}                & 40                           & 40                  & 400                    \\ \hline 
\textbf{Parallel overall time}     & 2.09 days                    & n/a                 & 4.31 days              \\ \hline 
\textbf{Observed time per job}     & 5.6 hours                    & n/a                 & 1.07 days              \\ \hline 
\textbf{Effective time per job}    & 8.40 minutes                 & n/a                 & 3.85 minutes           \\ \hline 
\textbf{HPC Efficiency}            & 22.30\%                      & n/a                 & 16.50\%                \\ \hline 
\textbf{Effective scale up factor} & 8.91                         & n/a                 & 65.6                   \\ \hline
\end{tabular}

 \caption{Metrics of grid performance compared with serial execution.  Note that Parallel overall time for the Quasar job was not recorded, and as such quantities derived from it cannot be calculated.  Copied from \citet{creaner2016thesis}}
 \label{table:gridmetrics} 
\end{table*}

\subsubsection{API Grid Considerations}
\label{APIGrid}
In the case of using the API in grid mode to generate the Local Catalogue, the SDSS Catalogue was divided into batches of 1000 input files, each batch therefore forming a grid job.  Each job was, as shown in Table 2, dominated by file I/O operations to the LFC.  Parallel file access to the LFC was faster than serial access from a single device, but did not provide linear scale-up with additional nodes added.  With 40 nodes used, an overall scale up factor of 8.91 was achieved.  This scale-up was calculated by comparing the observed mean time for serial execution of a single grid job of 1.25 hours multiplied by 422 (the total number of jobs) to give a serial execution time of 21.9 days.  By contrast, the execution of the job in parallel mode took 2.09 days.  Each job in the parallel execution was slower than the serial counterparts, with an average execution time estimated at 5.6 hours.  This gives a HPC efficiency of 22.3\%.
\subsubsection{Pipeline Considerations}
\label{PipGrid}
The use of the pipeline on the grid to generate the Exoplanet Catalogue was evaluated in comparison with a serial computing solution.  Serial operations for a small sample of jobs were measured to take an average of 4.25 hours per job.  Multiplying this by the total number of jobs gave an estimated serial time of 317 days.  By running on the grid, this time with 400 nodes, the whole process took 4.31 days, with each grid job taking an average of 1.07 days.  This gives an effective scale up factor of 65.6 and a grid efficiency of 16.5\%.  Grid performance metrics for the generation of the Quasar Catalogue were not captured.

\section{Discussion}
\label{Discussion}

Completion of this project highlighted a number of issues with the use of distributed computing systems, specifically \texttt{glite}, for astronomical operations.  In addition, the metrics created during the project highlight distinct use cases for systems of this nature which should be considered in future projects.  These issues and use cases are discussed below.
\paragraph{Access Requests}
\label{Access}
One major issue which arose during this project was the suitability of the \texttt{glite} LFC system to large volumes of simultaneous access requests.  Because the \texttt{glite} system is designed with data security in mind, it activates separate authentication processes for each file accessed from the LFC. \cite{glite} This means that each file takes a minimum time of the order of one second to access and download.  For typical grid problems where the data is contained in few large files, this issue is not significant.  In this case, where each job may access thousands of small files, this access time can become dominant.  The solution to this problem was to provide storage space at a shared network drive (NFS) which lacked the thorough security and authentication process of the LFC.  This allowed for faster data access for grid jobs using that system.
\paragraph{Log Files}
\label{Logs}
It is typical for grid job to produce log files which include the standard outputs.  These log files are typically small, of the order of kilobytes.  In this project, development versions of the system had produced verbose outputs to the screen containing much of the data from input files and the final resulting file.  If stored, this output would be comparable to or larger than the input data, amounting to many megabytes per grid job.  Log file storage systems were unable to manage this data volume and velocity, and as a result a silent mode was developed for the software which discarded much of the data which would otherwise have been sent to std.out and thus to log files.
\paragraph{Performance Comparison}
\label{Comparison}
While the software for this project was structurally divided into two components: the API and the Pipeline, these components are not fundamentally different in terms of how they used the grid: each read in the data from the LFC (Input), processed the data through some software (processing) and produced the output to the LFC (Output).  These steps are present in each of these and many other applications.  A distinct use case emerges however between the generation of the Local Catalogue and Quasar Catalogue on one side, and the Exoplanet Catalogue on the other.  

The former case is dominated by Data I/O operations especially when the \texttt{glite} LFC is accessed and requires authentication for each file accessed. The latter case is balanced between Data I/O and Processing.  Finally, one can imagine a case where Processing is dominant, as is commonly the case with HPC applications.  Since the same system (the pipeline) demonstrates different characteristics dependent upon the inputs it is provided with, it follows that a simple examination of the code or algorithm design may not always be indicative of the characteristics of the system overall.

In future applications of this or related systems, therefore, it becomes important to distinguish not only the characteristics of the system, for example by analysis of algorithmic complexity, but also to test those system characteristics when applied to data of the type that is to be processed.

\section{Conclusions}
\label{Conclusions}

The objective of this project was to generate a system to analyse and identify optimal pointings for differential photometry for two sets of targets: Quasars (from SDSS Quasar Data Release 4) and stars which might be potential candidates to host exoplanets.  In both cases, optimal conditions for differential photometry were defined such that the target had a maximum number and quality of reference stars as defined in \citet{locuspaper}

To achieve this objective, two major software components were developed, and used to generate three catalogues, the Local Catalogue, the Quasar Catalogue and the Exoplanet Catalogue.  The Local Catalogue was used as an intermediate step in the generation of the two Output Catalogues.  Generation of these catalogues was completed using the \texttt{glite} system at the National Grid Infrastructure managed by Grid-Ireland.  

Operational metrics on these systems showed a disparity in performance not between the software systems but rather between the application of the systems to different datasets.  The ratio of file I/O to data processing time was heavily dominated by file I/O in the case of the API and Quasar Catalogue grid jobs, while File I/O and processing were approximately balanced in the case of the job to generate the list of targets for an Exoplanet search. This was despite the fact that, from an algorithmic perspective, the Exoplanet and Quasar Catalogue jobs use the same software, and thus would be predicted to be an operation of the same order.  It can also be seen that in a case where the data is accessed even more efficiently, the processing time could come to dominate. 

From these metrics and practical observations, it is apparent that simple calculations of algorithmic complexity may not be sufficient to identify the dominant component in a project.  Rather, this system demonstrates the value of experimental assessment of system performance under operational conditions.

\section*{Acknowledgements}
\textbf{Funding for this work}: This publication has received funding from Higher Education Authority Technological Sector Research Fund and the Institute of Technology, Tallaght, Dublin Continuation Fund (now Tallaght Campus, Technological University Dublin).

\textbf{SDSS Acknowledgement}: This paper makes use of data from the Sloan Digital Sky Survey (SDSS).  Funding for the SDSS and SDSS-II has been provided by the Alfred P. Sloan Foundation, the Participating Institutions, the National Science Foundation, the U.S. Department of Energy, the National Aeronautics and Space Administration, the Japanese Monbukagakusho, the Max Planck Society, and the Higher Education Funding Council for England. The SDSS Web Site is http://www.sdss.org/.

The SDSS is managed by the Astrophysical Research Consortium for the Participating Institutions. The Participating Institutions are the American Museum of Natural History, Astrophysical Institute Potsdam, University of Basel, University of Cambridge, Case Western Reserve University, University of Chicago, Drexel University, Fermilab, the Institute for Advanced Study, the Japan Participation Group, Johns Hopkins University, the Joint Institute for Nuclear Astrophysics, the Kavli Institute for Particle Astrophysics and Cosmology, the Korean Scientist Group, the Chinese Academy of Sciences (LAMOST), Los Alamos National Laboratory, the Max-Planck-Institute for Astronomy (MPIA), the Max-Planck-Institute for Astrophysics (MPA), New Mexico State University, Ohio State University, University of Pittsburgh, University of Portsmouth, Princeton University, the United States Naval Observatory, and the University of Washington.





\bibliographystyle{elsarticle-harv}
\bibliography{grid_metrics}







\end{document}